\documentclass[amssymb,prb,aps,showpacs,twocolumn,showpacs]{revtex4}
\usepackage{graphicx}
\usepackage{amsmath}
\usepackage{epstopdf}
\usepackage{citesort}
\begin{document}

\title{\textbf{Ultrafast quantum spin-state switching in the Co-octaethylporphyrin
molecular magnet with a terahertz pulsed magnetic field}}
\author{Oleg V.Farberovich$^{1,2,3}$,Victoria L.Mazalova$^{2}$, Valeri S.Stepanyuk$^{4}$}
\affiliation{$^1$School of Physics and Astronomy, Beverly and Raymond Sackler Faculty of Exact Sciences,
Tel Aviv University, Tel Aviv 69978, Israel\\
$^2$International Center "Smart Materials", Southern Federal University,
Zorge 5, 344090, Rostov-on-Don,Russia\\
$^3$Voronezh State University, Voronezh 394000, Russia\\
$^{4}$ Max Planck Institute of Microstructure Physics, Halle, Germany\\}
\date{\today}

\begin{abstract}
Molecular spin crossover switches are the objects of intense theoretical and
experimental studies in recent years. This interest is due to the fact that
these systems allow one to control their spin state by applying an external
photo-, thermo-, piezo-, or magnetic stimuli. The greatest amount of research
is currently devoted to the study the effect of the photoexcitation on the
bi-stable states of spin crossover single molecular magnets (SMMs).
The main limitation of photo-induced bi-stable states is their short lifetime.
In this paper we present the results of a study of the spin dynamics of
the CoOEP molecule in the Low Spin (LS) state and the High Spin (HS)
state induced by applying the magnetic pulse of $36.8 T$. We show that
the spin switching in case of the HS state of the CoOEP molecule is characterized
by a long lifetime and is dependent on the magnitude and duration of the applied field. Thus, after applying an external stimuli the system in the LS state after the spin switching reverts to its ground state, whereas the system in the HS state remains in the excited state for a long time. We found that the temperature dependency of magnetic susceptibility shows an abrupt thermal spin transition between two spin states at 40K. The proposed here theoretical approach opens the way to create modern devices for spintronics with the controllable spin switching process.
\end{abstract}

\maketitle
\section{Introduction}
Spin engineering mostly concentrates on the manipulation of
spin degrees of freedom in various magnetic materials and
the exploration of their potential applications in spintronics \cite{i8}.
Current examples for the applications of spin engineering
in magnetic nanostructures and devices are increasingly
ample, ranging from the development of single molecular magnet
as qubits \cite{1}. Therefore, the spin engineering in magnetic nanostructures,
including theoretical modeling, experimental demonstration,
and device design for applications, is demanding multidisciplinary
backgrounds of knowledge and technology which create great challenges for researchers.

Magnetic or spin logic appears as an appealing alternative
due to its nonvolatile character, which can boost up
switching on/off, its possibility to reduce the size of the element
down to the several-atoms scale on/off spin per atom
instead of one elementary charge per $10^4$ atoms in semiconductors,
and speed increase as a secondary size effect \cite{2}.
The modern experiments are promising, they still move in the micrometer
regime \cite{3}, thus not fully exploiting the possible quantum
nature of molecular magnetism. In other hand, small molecules perform logic operations
using as input cation concentrations. The latter is fast with
respect to the logic operation but slow with respect to repeatability.
Thus a need for magnetic-logic devices on the molecular
scale emerges. At the same time the experimental
evidence of laser-driven ultrafast magnetic(spin) manipulation in
the antiferromagnetic materials motivates the design of a
cluster with more than one spin center which allows for
spin manipulation \cite{4,5}.

The problem of the gigahertz magnetization(spin) switching
speed of today’s magnetic logic and a magnetic
memory devices into the terahertz regime
underlies the entire field of information
processing \cite{6,7}. The physics of present-day devices
imposes serious limitations on this technological
transformation, so we must invent new
paradigms based on the $quantum$ $spin$ $dynamics$ in the
picosecond regime. This challenge could be
met by a simulation of the quantum spin switching
in a picosecond pulsed magnetic field \cite{W27,F8}.
Currently, no easily accessible method is available
to generate intense sub-picosecond magnetic pulses localized at the nanoscale.
However in the work \cite{9}, show that bimetallic nanorings can act as nanoscale
sources of intense ultrashort magnetic pulses.
In this instance rely on the enhanced light absorption associated with the plasmons
of metallic rings to generate transient thermoelectric
currents that in turn produce sub-picosecond pulses with magnetic
fields as high as a few tenths of a Tesla
in the vicinity of the rings. The ability to generate strong
magnetic fields localized on the nanoscale is of interest for
elucidating spin and magnetization dynamics at sub-picosecond
time and nanometer length scales \cite{10}, and it holds
great potential for materials characterization, terahertz
radiation generation, and magnetic recording.

Magnetic molecules have been intensively studied so
far because of their potential technological
applications and for
the possibility they offer to investigate fundamental
properties
of matter at the nanoscopic scale \cite{11}.
An intensively studied class of magnetic molecules
is constituted by the so called single molecule magnets or molecular
nanomagnets. These systems are characterized by
a slow relaxation of magnetization
at low $T$ and give rise to magnetic hysteresis.
This is one condition for storing
information in a molecule. Therefore, since a SMM represents
an isolated magnetic
entity, it might be considered as the smallest practical unit
for magnetic memories.
The studying of a possibility of the switching of the spin states
would incredibly increase the amount of information storable with
respect to
current devices. In order to have a deeper understanding
of fundamental features and a major control on technological aspects, the
time-dependent spin dynamics constitutes the key point.
The micromagnetic model describes appropriately the dynamics of the
magnetization(spin) in a magnetic nanostructures. In this model
the magnetization(spin) is
assumed to be a spatial- and time-dependent continuous function.
The magnetization(spin) dynamics are described by the
Landau-Lifshitz-Gilbert (LLG)
equation including the energy contributions
of the anisotropy, the exchange interaction, the magnetostatic
interaction, and the Zeeman
energy \cite{LLG}.
The microscopic result of the quantum modeling  is the
cluster spin models, which have been proved to be a
powerful tool for approaching the extreme phenomenology
of the ultrafast magnetization(spin) dynamics.
 Ab initio models for the general spin Hamiltonian can be
efficient in providing insight on the local atomic scale values
such as the local magnetic moment $\mu_s$, the local anisotropy
$D$, or a pairwise exchange $J_{ij}$ in the magnetic molecule.
The parameters of the theory, the exchange integrals
and the anisotropy, are usually fitted to experiments or
calculated from static density functional theory (DFT) \cite{12}.

Switching between the LS and HS states can be done by
varying the temperature, applying pressure or by light
irradiation. In the first and the second case the population of
the LS and HS states is fully determined by thermodynamic
equilibrium and such transitions are called temperature- and
pressure-induced spin crossover transitions \cite{13}. In the case of
light-induced spin crossover, after initial excitation to d-states
or to metal-to-ligand charge transfer states, the system undergoes
a complex cascade of intermediate transitions and eventually
gets trapped in a non-equilibrium metastable state. The
whole process is called light-induced excited spin state trapping
(LIESST). The details of the LIESST process are still under
debate \cite{14}.

We report here the possibility of a commutation
of the CoOEP SMM between the two states of a spin-crossover system by
a pulsed magnetic field \cite{15}. Choosing a cobalt-based spin-crossover
system follows the idea that cobalt(II)[CoOEP] materials \cite{16} as well as
iron(III) ones have metal-ligand bonds shorter than those in
iron(II) materials, leading to faster relaxation dynamics between
the HS and LS states. The molecule cobalt(II) ([Ar]3$d^7$4$s^2$)
may be either in the LS state
($\uparrow\downarrow\uparrow\downarrow\uparrow\downarrow\uparrow S=1/2$)
or in the HS state ($\uparrow\downarrow\uparrow\downarrow\uparrow\uparrow\uparrow S=3/2$),
both of which are paramagnetic.

\section{THEORETICAL ASPECTS}

\subsection{The spin-dynamics simulations}
The dynamic behavior of a spin is determined by the equation
of motion, which
can be derived from the quantum theory with the general spin Hamiltonian
$\widehat H_{spin}$ that calculated the spin structure of a magnetic molecule with
Hamiltonian
\begin{equation}\label{GSH}                                                                                                                 \widehat H_{spin} = \widehat H_{ex} + \widehat H_{an} +
\widehat H_{ZEE} + \widehat H_a(t).
\end{equation}                                                                                                                   The first term $\hat H_{ex}$ is the Heisenberg-Dirac Hamiltonian, which represents the isotropic
exchange interaction, $H_{an}$ is the exchange Hamiltonian the term due to the
axial single-ion anisotropy, and $H_{ZEE}$ is the interaction between the
spin system and the external magnetic field.
In order to give a theoretical description of a magnetic molecule
we exploit the irreducible
tensor operator technique \cite{ITO20}.
For the study of the CoOEP magnetic properties,
it is a good approximation to treat the electrons of one Co-atom as a single atomic
spin which is interacting with its surroundings. Thus, one can be content
with $\widehat H_{spin}$ without the first term  $\hat H_{ex}$.
We will consider an approximate treatment, restricting ourselves to the simplest,
but very useful, approximation, known as
mean field theory.
This "mean field" Hamiltonian \cite{16}, which describes
the interaction of the spin
$\widehat{\bf S}$ with the external
magnetic field, given by its flux ${\bf H}_{eff}$, can be expressed as:
\begin{equation}
\widehat H_{spin}=-\gamma{\bf H}_{eff}\widehat{\bf S}.
\end{equation}
The generally shaped formula is
\begin{equation}
{\bf H}_{eff}=-\frac{\partial\widehat H_{spin}}{\partial\langle\widehat {\bf S}\rangle}.
\end{equation}
Here the effective magnetic field ${\bf H}_{eff}$ is
an external magnetic
field $H_z$, the anisotropy
fields ${\bf H}_{an}$, the exchange interaction ${\bf H}_{ex}$
and external magnetic pulse field ${\bf H}_{pulse}(t)$.
We use the "mean field" \cite{16} approximation for ${\bf H}_{eff}
\Rightarrow{\bf H}_{eff}^{mean}$
with replacement $\widehat{\bf S}\Rightarrow{\bf M}_s=\gamma\langle\widehat
{\bf S}\rangle$.
We call "mean" the mean field. (Historically it was often called the "molecular field"). It includes
the average effect of the neighbors but neglects correlations between the spin and its neighbors.
Since we have the one magnetic atom, this approach us quite suitable.
Use (2) we obtain that \cite{17}
\begin{equation}
\frac{\partial\langle\widehat {\bf S}\rangle}{\partial t}=\frac{1}{1+\lambda^2}
\langle\widehat {\bf S}\rangle\times{\bf H}^{mean}_{eff}-
\frac{\lambda}{1+\lambda^2}\langle\widehat{\bf S}\rangle\times
(\langle\widehat{\bf S}\rangle\times{\bf H}^{mean}_{eff})
\end{equation}
The effective magnetic field ${\bf H}_{eff}^{mean}$ is given by the
free magnetic energy
variational with magnetization:
$$                                                                                                                                                               {\bf H}_{eff}^{mean}({\bf M}_s,t) =-\frac{\delta F}{\delta{\bf M}_s},
$$                                                                                                                                                                where $F$ is the free energy of the magnetic nanosystem                                                                                               $$                                                                                                                                                               F=-Nk_BT\ln Z(H_z)                                                                                                                                    $$                                                                                                                                                               with the partition function
\begin{equation}\label{Z}                                                                                                                                                 Z(H_z) = \sum_{M_s,\mu}\exp[-\epsilon_\mu(M_s)/kT]
\sum_{M_s}\exp[-g_e M_s H_z/kT]
\end{equation}.
Here we have the energy levels $\epsilon_{\mu}(M_s)$
of the spin-Hamiltonian $\widehat H_{spin}$, which stay we diagonalizied
$\widehat H_{spin}$.

A spin structure is defined only proceeding from the spin model of a molecule.
Here we use to calculate a spin structure by the ITO method within the generalized spin
Hamiltonian $\widehat H_{spin}$.

Once we have the energy levels of the spin-Hamiltonian, we can evaluate a
different thermodynamic properties of the system as the magnetization,
the magnetic susceptibility, and the magnetic specific heat.
Since in  further researches the anisotropic part of a molecule will be only scalar,
the magnetic properties of the anisotropic system do not depend on the
direction of the magnetic field.
Thus we can consider the external magnetic
field $H_z$ directed along arbitrary axis
$\it z$ of the molecule coordinate frame that is
chosen as a spin quantization axis. In
this case the energies of the system will be
$\epsilon_{\mu}(M_s) +   g_e M_s H_z$, where
$\epsilon_{\mu}(M_s)$
are the eigenvalues of the spin-Hamiltonian containing
the magnetic exchange and
the double exchange contributions
(index $\mu$ runs over the energy levels with given
total spin protection ${\bf M}_s$).
Using this expression one can evaluate the magnetization
${\bf M}_s$ by the standart thermodinamic definitions:
\begin{equation}\label{M}                                                                                                                                                 {\bf M}_s=\frac{\partial F(M_s,H_z)}{\partial{\bf H}}=
NkT\frac{\partial{\ln Z}}{\partial \bf H}                                                                                                     \end{equation}

The effective field ${\bf H}_{eff}^{mean}$ can be derived from
the free energy functional
\begin{equation}
{\bf H}_{eff}^{mean}=-\frac{\delta (F(M_s,H_z)+F(t))}{\delta{\bf M}_s}=
-\frac{\partial F(M_s,H_z)}{\partial{\bf M}_s}+H_{pulse}^x(t),
\end{equation}

We have derived a general form of the time-dependend spin equation for a system of the spins                                                                             precessing in an effective magnetic field with specifying the interactions
in the magnetic molecule.

\subsection{Calculations of spin-Hamiltonian parameters}

The present research is devoted to study a time-dependent behavior
of the Co-octaethylporphyrin molecule after applying an external
magnetic field in a picoseconds time span. In order to optimize
the molecule geometry and determinate the spin-Hamiltonian
parameters - zero-field splitting (ZFS, D-tensor), g-tensor,
the exchange parameters $J_{ij}$ , and the single-ion anisotropy parameters,
we performed Density Functional Theory (DFT) calculations with taking into
account the spin-orbit coupling effects using ADF package \cite{ADF, ADFF}.
The calculations were performed with the Perdew-Burke-Ernzerhof (PBE)
exchange-correlation functional and the ZORA scalar relativistic Hamiltonian,
using an all-electron valence triple zeta with a one polarization function
(TZP) basis set and a 'Good Becke' grid. The zero-field splitting (ZFS)
of the Co OEP ground state was found to be equal 1.32meV. ZFS is the breaking of degeneracy
of the ground state that is not described by a standard non-relativistic
Hamiltonian. As calculated by ADF, the ZFS is that exhibited by molecules
whose ground state characterized by the spin S$>$1/2 and absence of a spatial
degeneracy. This type of ZFS has two contributions, second-order spin-orbit
coupling and spin-spin coupling. In the present implementation only the
spin-orbit coupling term is included. In our study the ZFS was calculated
in combination with GGA PBE functional. In order to calculate ZFS the
relativistic scalar Zero Order Regular Approximation (ZORA) option
\cite{ZORA} was included.
The ZFS Hamiltonian is given by
\begin{equation}
\widehat H_{ZFS} = \widehat{\bf S}\cdot{\bf D}\cdot\widehat{\bf S}
\end{equation}

where $\widehat{\bf S}$  is the vector operator for the effective spin of Co atom, and D is the second-order anisotropy tensor. After diagonalization in the magnetic axis frames, the Hamiltonian becomes

\begin{equation}\label{ZFS}
\widehat H_{ZFS} = D\widehat S_{z}^2 + E(\widehat S_{x}^2 - \widehat S_{y}^2)
\end{equation}

The parameter D is called the axial ZFS parameter and E is the rhombic
ZFS parameter, which vanishes for high symmetry molecules. Parameter D
has to be negative for the magnetic bistability assosiated with SMMs.
When a molecule is placed in a magnetic field ${\bf H}$,
the electron's energies will depend on its magnetic moment value ${\bf M}_s$.
To account for the anisotropy of the Zeeman response to an applied magnetic
field, an "effective" Zeeman Hamiltonian using a so-called "g tensor" is used
\begin{equation}
{\bf\widehat H}_{ZEE} = \mu_{B}{\bf H}\cdot{\bf g}\cdot{\bf\widehat S} = \mu_{B}
\begin{bmatrix}
H_{x}&H_{y}&H_{z}
\end{bmatrix}
\begin{bmatrix}
g_{x}&0&0 \\
0&g_{yy}&0 \\
0&0&g_{zz}
\end{bmatrix}
\begin{bmatrix}
\widehat S_{x} \\
\widehat S_{y} \\
\widehat S_{z}
\end{bmatrix}
\end{equation}
The g-tensor was calculated in a spin-orbit coupled spin unrestricted relativistic ZORA approach. The degenerate perturbation theory was used with the external magnetic field as perturbation.
\section{Results and discussion}
The most important aspect to the establishment of real devices based on SMMs,
e.g. high density rewritable memories, is to achieve such conditions under that
externally driven system could be switched between two stable LS and HS
spin states and would remain in each of these states for a certain period of time.
The LS/HS bi-stability is connected with small energy
($\Delta E_{HS-LS}$=2.133 meV ) magnitudes
involved in the switching between these two spin states (Figure 1a).
Much research
is currently focused on the study of photoswitching mechanism between LS and
HS states, changes in the geometric and electronic structure of molecules in
these states, the magnetic properties of SMMs. However, we found no studies
on the time-dependent spin dynamics of spin-crossover systems.
\begin{figure}[tbp!]
\begin{center}
\includegraphics[width=1.0\columnwidth]{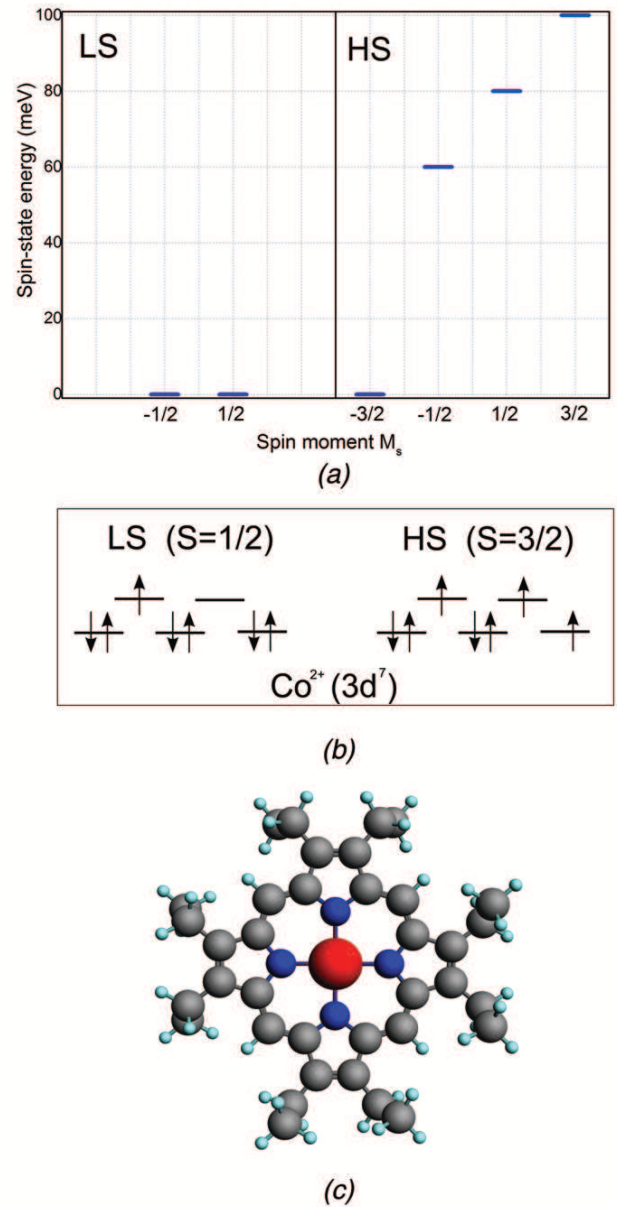}
\end{center}
\caption{(Color online) (a)Spin structure of CoOEP molecule in a spin crossover
system HS-LS states;(b)HS and LS states with $3d^7$ configuration of $Co^{II}$
in a tetrahedral field;(c)The molecular structure of CoOEP.}\label{fig.1}
\end{figure}
In this paper
we focus on the behavior of the spin of the molecules under the influence
of external stimuli and finding the conditions under which the system
would be able to stay in one of the bistable states arbitrarily long time.
Since the photo-induced states of SMMs are characterized by a short lifetime,
we effect on the system being studied by a short magnetic pulse of high amplitude.
In the experimental study \cite{15}, it was shown that the application of
an external magnetic field of $32 T$ to spin-crossover solid system
$Co(H_{2}(fsa)_{2}en)(py)_{2}$ takes place an irreversible and quasicomplete
transition from the HS to the LS state. This also means that after applying
a perturbation the system can remain in HS state for a long time.
In this paper,
we have developed an approach for the study of the spin dynamics of spin-crossover
systems and apply it to one of the intensively studied objects from a number
of SMMs – the CoOEP molecule. The chemical structure of CoOEP molecule is
exhibited in Fig. \ref{fig.1}(c).
Fig.\ref{fig.2} shows the magnetic susceptibility plotted as the product $\chi T$
versus temperature for the CoOEP molecule. The temperature variation of the
magnetic susceptibility $\chi T$ showing the abrupt of the thermal spin
transition with the jump between two spin states occurs at temperature of
$T_{1/2}$=40K.
\begin{figure}[tbp!]
\begin{center}
\includegraphics[width=1.0\columnwidth]{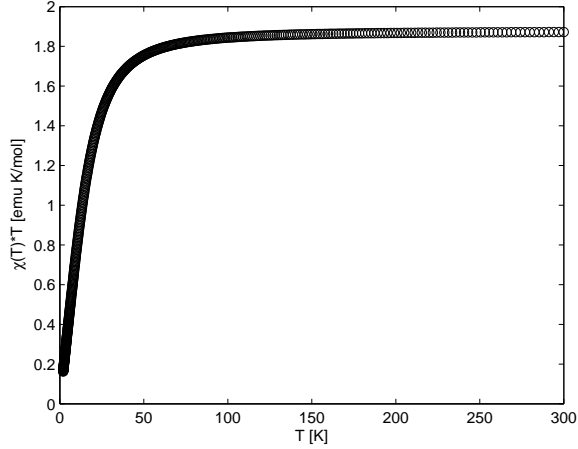}
\end{center}
\caption{(Color online) The magnetic susceptibility product $\chi$ T.} \label{fig.2}
\end{figure}

By irradiating LS molecules with visible light, HS molecule can be generated because of the
efficient electronic decay of excited state via Inter System Crossing towards
matastable HS state. This effect is now referred as LIESST (Light Induced
Electronic Spin State Trapping) in the literature \cite{L}. Once
irradiation is stopped, the system relaxes to the thermal equilibrium state LS.
One important LIESST-result  is the decoupling of different degrees of freedom
during the electronic spin state trapping and the transfer of excess energy
to molecular vibration modes. Contrary to the LIESST, here considered in some
detail the quantum dynamics of the HS spin state trapping and impacts
on it of a short magnetic pulse.

\begin{figure}[tbp!]
\begin{center}
\includegraphics[width=1.0\columnwidth]{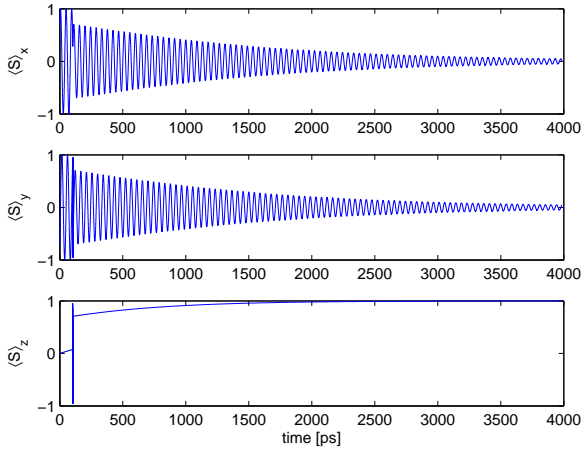}
\end{center}
\caption{(Color online) The spin dynamics for the CoOEP molecule
with magnetic pulse (S=$1/2$).} \label{fig.5}
\end{figure}

We present here a theoretical approach to the organic
magnetic CoOEP-molecule in terms of numerically
solvable the time-dependent LLG-model with a general
spin Hamiltonian in the context of the ITO model.
In practice, these models apply to the
quantum spin-state switching in the Co-octaethylporphyrin
molecular magnet with a picosecond pulsed magnetic field and provide
an understanding of a spin-crossover phenomenon in these molecule.
In the calculations by LLG-method was used the next parameters:
$T_{pulse}$=0.17$ps$ (ultrashort THz pulse); time of start pulse is 37.4$ps$; the height
of pulse is 36.8$T$.
We solved the LLG equation for two different states S=1/2, $t^{6}_{2g}e^1_{g}$ (LS)
and S=3/2, $t^5_{2g}e^2_g$ (HS). Under the influence of the constant
Zeeman magnetic field 0.1$T$, after a certain period of time, which is
determined in case of a single molecule by the size of ZFS, the time-dependent
spin structure is formed. In the time of "saturation" of the system we apply
the external stimuli of a magnetic pulse. The system in the LS state
behaves classically without the application of the pulse.
Under the influence of the 36.8$T$ pulse a spin switching on the pulse duration
of $100 fs$ is happened, after which the system returns to the ground state
(Fig.\ref{fig.5}).
In the case of the HS state of the system the situation is
changing drastically. Under the influence of the pulse a spin switching
occurs as well, but the system remains in this state for a long time (Fig.\ref{fig.6}).
\begin{figure}[tbp!]
\begin{center}
\includegraphics[width=1.0\columnwidth]{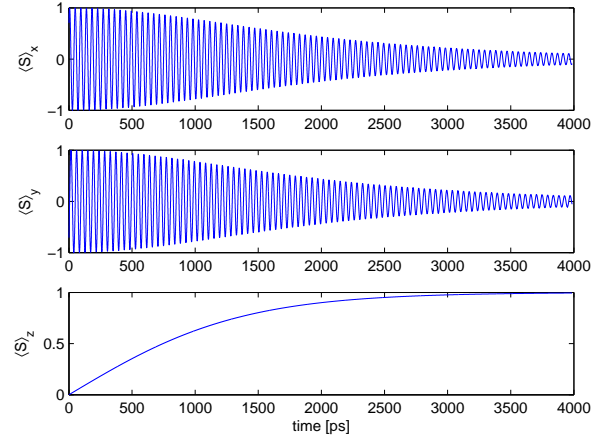}
\end{center}
\caption{(Color online) The spin dynamics for the CoOEP molecule
without magnetic pulse(S=$1/2$)} \label{fig.3}
\end{figure}
With increasing magnetic field up to 36.8$T$ a spin switch
occurs and the $H_{eff}$ sign is changed.
In our approach $H_{eff}$ is the $\langle\widehat{\bf S}\rangle$ derivative
of the spin-Hamiltonian. In our case, the Hamiltonian (\ref{GSH}) has no
$H_{ex}$ term, because there is only one transition metal atom in CoOEP molecule and,
accordingly, the spin-spin interaction is absent. Thus, the anisotropic $H_{an}$
term is only a nonconstant one, which is quadratic in
$\langle\widehat{\bf S}\rangle$ (\ref{ZFS}) and, therefore, $H_{eff}$ is
proportional to $H_{eff}$ is the $\langle\widehat{\bf S}\rangle$.
Qualitatively, the $H_{eff}$ behavior is the same as the  $\langle\widehat{\bf S}\rangle$,
which is seeing from the Fig .\ref{fig.4} and \ref{fig.6}.
\begin{figure}[tbp!]
\begin{center}
\includegraphics[width=1.0\columnwidth]{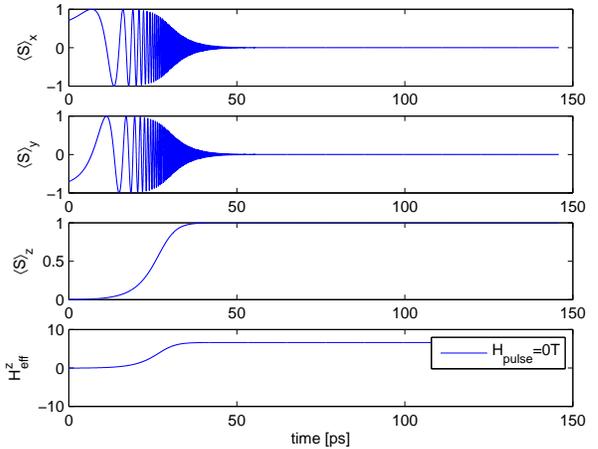}
\end{center}
\caption{(Color online) The spin dynamics for CoOEP molecule
without magnetic pulse (S=$3/2$).} \label{fig.4}
\end{figure}
\begin{figure}[tbp!]
\begin{center}
\includegraphics[width=1.0\columnwidth]{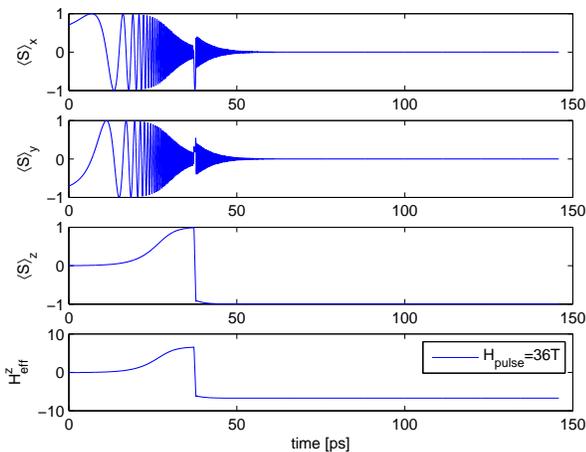}
\end{center}
\caption{(Color online) The spin dynamics for the CoOEP molecule
with magnetic pulse (S=$3/2$).} \label{fig.6}
\end{figure}
%
%
%
%
\section{Concluding remarks}
In the paper we present the results of the theoretical study of the spin behavior in the CoOEP molecule under influence of high magnetic pulse. Based on the LLG equation we described the spin dynamics of the molecule in the LS and HS states. We found that the spin-crossover CoOEP molecule undergo the spin switching between LS and HS states after applying the external stimuli of a short magnetic pulse of 36.8T. The temperature dependency of magnetic susceptibility shows an abrupt thermal spin transition between two spin states at 40K. We also found the difference in the spin dynamics for two spin states of CoOEP molecule. Thus, after the spin switching under the magnetic pulse the system in the LS state reverts to its ground state, whereas the system in the HS state after the spin switching remains in the excited state for a long time. This behavior gives us the possibility to assign such system as an appropriate candidate for the SMM and opens a way to realize a controlled spin switching in real devices for spintronics.

\begin{acknowledgements}
The part of this research has been supported by Russian Ministry of Science, Grant No. 213.01-11/2014-6 (for V.L.M.)  and RFMEFI 58714X0002 (for O.V.F.). The work is partially supported by the Max-Planck Gesellschaft during the stay of O.V.F. in MPI (Halle) where this work was completed. The authors are indebted to Prof. Alexander Soldatov and Dr. Alexander Guda for detailed discussion.
\end{acknowledgements}

\end{document}